\documentstyle[11pt,epsfig]{article} 
\def\baselinestretch{1.3}
\newcommand{\ba}{\begin{array}}
\newcommand{\ea}{\end{array}}
\newcommand{\bd}{\begin{displaymath}}
\newcommand{\ed}{\end{displaymath}}
\newcommand{\be}{\begin{equation}}
\newcommand{\ee}{\end{equation}}
\newcommand{\bea}{\begin{eqnarray}}
\newcommand{\eea}{\end{eqnarray}}

\def\q2 {q^2}

\begin{document}

\begin{center}
{\Large\bf Supersymmetry and neutrino mass}\\[20mm]
Biswarup Mukhopadhyaya\footnote{E-mail: biswarup@mri.ernet.in}\\
{\em Harish-Chandra Research Institute,\\
Chhatnag Road, Jhusi, Allahabad - 211 019, India}, 
\\[20mm] 
\end{center}
The existence of neutrino mass and mixing is a strong pointer towards physics
beyond the standard model.
An overview of the possibility of having neutrino masses in supersymmetric 
theories is attempted here. Some of the recent works reviewed suggest 
Dirac masses, whereas others include Majorana masses as well. Side by side,
it is shown how R-parity violating supersymmetry opens new avenues in the
neutrino sector. Reference is also made to light sterile neutrinos, 
nearly degenerate neutrinos and neutrinos acquiring masses from 
hard supersymmetry breaking terms which are suppressed by the Planck scale.
In several of the cases, it is pointed out how the models that give
neutrino masses and mixing have independent motivations of their own,
and can be tested in accelerator experiments. 

\vskip 5 true cm
\begin{center}
{\it To appear in the Proceedings of the Indian National Science Academy, 
special volume on Neutrinos}
\end{center}

\newpage
\setcounter{footnote}{0}

\def\baselinestretch{1.8}

\section{Introduction}

As has been amply established in the other articles in this volume, there is
a strong evidence nowadays in favour of neutrino masses. In addition,
the solar \cite{sol} and atmospheric \cite{atm} neutrino data 
have their most obvious 
explanation in neutrino oscillation, requiring mixing among neutrinos,
or, more generally speaking, in the leptonic sector, in analogy with quark
mixing which is controlled by the 
Cabibbo-Kobayashi-Maskawa (CKM) matrix \cite{km}. However, in
contrast to quark mixing, the most favoured explanations of the solar and
atmospheric neutrino deficits require very large-- even close to maximal--
mixing between the first two families and the last two. Side by side,
the data indicate a hierarchy of mass splitting, the mass-squared difference
being in the range $10^{-3}~-~-10^{-2}~eV^2$ between the second and
the third families, and, most favourably,  $10^{-5}~-~-10^{-4}~eV^2$
between the first and the second. Though such splittings are most often
translated into a corresponding hierarchy in the masses themselves, the
existence of near-degenerate neutrinos, too, cannot be ruled out.

According to many, all this is an indication of physics beyond the standard
model. To see why, let us recall that, thanks to the electrically neutral 
character of neutrinos, they can have both Dirac and Majorana masses. While
the second possibility which entails lepton number violation clearly entails 
new physics, albeit at high scale, the first one can be {\it prima facie}
dismissed as a `trivial' extension in the form of a right-handed neutrino
component for each family. However, the fact that such a right-handed neutrino
has none of the strong, weak and electromagnetic interactions
is curious, if not suggestive of some new interaction in which takes part.
The extreme suppression of neutrino Yukawa couplings necessitated by
sub-$eV$ Dirac masses is also puzzling. Side by side, if the LSND claim 
suggesting the disappearance of $\nu_{\mu}$'s is to taken seriously cite{lsnd},
we most likely need a fourth light neutrino, sterile in nature. Since the
mass of a sterile vectorlike neutrino is not protected by any symmetry,
and since we can hardly think of any new physics scale below a $TeV$ or so,
a light sterile neutrino, if it is there at all, warrants a drastically
novel mechanism for its justification.

The new physics scale to which appeal has mostly been made to understand
neutrino masses is that pertaining to Grand Unified Theories (GUT),
restricted to be at least about $10^{16}~GeV$ \cite{gut}. 
However, there are other
motivations for physics beyond the standard model within the TeV scale itself.
One cause such expectation is
the so-called naturalness problem which reflects our lack of understanding why
the Higgs mass (and consequently the electroweak scale $M_{EW}$) should be
stable against quadratically divergent radiative corrections. The most 
popular solution to this problem has been offered in terms of supersymmetry 
(SUSY), a symmetry between bosons and fermions, which can provide the 
necessary cancellations to control the large 
radiative corrections \cite{susy1}. Most 
importantly, it is possible to keep the Higgs mass within acceptable limits
even if SUSY is broken {\it in mass}, so long as the breaking scale
(characterising the boson-fermion splitting) is approximately within the 
TeV scale. Side by side,
the observation that the threshold effects arising from Tev scale SUSY 
breaking ensures better convergence of the three coupling 
constants at the GUT scale provides an added impetus to SUSY \cite{susygut}.

In the minimal SUSY standard model (MSSM) \cite{mssm}, the 
particle spectrum of the standard model (SM) gets doubled, 
there being a superpartner for each known particle, apart from
the necessity of two Higgs doublets which lead to three neutral and a pair of
mutually conjugate singly charged scalars. There is no experimental
evidence yet for any of these superparticles; collider experiments have set
lower bounds of about 100 GeV upwards on most of them. Further consequences 
of SUSY also depend on the details of the spectrum which in turn is 
crucially dependent on the SUSY breaking mechanism. We know that
SUSY has to be broken at any rate if it is there, since we do not observe 
degenerate superpartners for the SM particles. No completely acceptable SUSY 
breaking scheme has been found so far, although most studies depend upon
a scenario based on $N=1$ supergravity (SUGRA) \cite{sugra}
where gravitational interactions 
with a  `hidden sector' characterised by a high scale 
($O(\sqrt{M_{P}M_{EW}})$) leads to soft SUSY breaking terms in the observable 
sector. In addition, schemes of SUSY breaking, for example, via gauge 
interactions of a messenger sector \cite{gmsb} or via 
anomaly terms \cite{amsb} have also been investigated.

The question is: since the search for physics beyond the standard model
has found a strong candidate in SUSY, could SUSY also be responsible
for neutrino masses (and mixing), the clue that nature seems to dangle
so tantalisingly in front of us? If that be so, then the
mass patterns answering to the solar and atmospheric neutrino data
should not only  depend on certain specific aspects of the SUSY model,
but also impose constraints on it. It may also be more 
convincing if models are built not just to answer questions on neutrinos
but have independent motivations of their own from the viewpoint 
of SUSY as well. Side by side, since the the search 
for SUSY is already an important goal of accelerator experiments,
it should be really interesting to look for the particular  
signatures of  such theoretical schemes as are able explain 
the observations in the neutrino sector. In other words, the issue of
neutrino masses could provide not only useful guidelines for theorisation,
but might also end up predicting specific experimental signals in
high-energy colliders. The present article is aimed at discussing some of 
these possibilities.

In very general terms, some of the ways in which SUSY can be of special
significance to neutrino masses are as follows: 

\begin{itemize}
\item The phenomenon of SUSY can provide new scales (in addition to that
brought by GUT in which most SUSY theories are embedded). These scales
open up additional possibilities in the neutrino sector and can be
helpful in explaining mass hierarchies. Also, some features of the 
SUSY theory might help us in understanding ultra-small Yukawa couplings.

\item The extended particle spectrum in SUSY can lead to mechanisms
for mass generation, for example, through additional radiative effects.

\item The possibility of low-energy 
lepton number violation inbuilt in certain types of 
SUSY theories might lead to the generation of Majorana masses.

\item SUSY could explain a naturally light sterile neutrino, 
in case we need it to explain the observed data.
\end{itemize}

In section 2 we discuss Dirac masses in presence of SUSY. Section 3 is
devoted to Majorana neutrinos in SUSY scenarios, where lepton number 
violation takes place at high-scale. Section 4 contains a summary of 
neutrino mass generation mechanisms in R-parity violating SUSY where 
the low-energy 
Lagrangian has lepton number violation. In section 5 we discuss
respectively the issues of degenerate neutrinos in SUSY and neutrino
masses from unusual SUSY breaking terms. We conclude in section 6.

\section{When lepton number is conserved--Dirac masses in SUSY}

If one takes the hierarchy in neutrino mass splitting to be an indication of 
the hierarchy in the masses themselves, then, assuming that the solar
and atmospheric neutrino deficits are due to $\nu_\mu  - \nu_\tau$
and $\nu_e  - \nu_\mu$ oscillations respectively, the two heaviest
neutrinos are about 10 to 11 orders of magnitude smaller in mass 
than the $\tau$ and the $\mu$. 
The simplest extension of the standard model spectrum that explains
the above masses is one right-handed neutrino per generation. However, the 
onus then falls on us to explain the wide disparity of Yukawa couplings
that is responsible for the huge mass splitting within the same families,
as indicated above. The question is: can SUSY provide some explanation
of such disparity?

Normally, with right-handed neutrino superfield $N$, one would expect a term 
in the superpotential of the form

\begin{equation}
W_N~=~y_{\nu} {\bar{N}}LH_2
\end{equation}

\noindent
where $H_2$ is the Higgs doublet giving mass to fermions with  
$T_3~=~1/2$. Of course, here one would find it hard to justify 
the smallness of $y_{\nu}$. On the 
other hand, one can forbid such a term with the help of  some discrete 
symmetry $Z_n$, and assume instead a higher-dimensional term \cite{borz1}

\begin{equation}
W_N~=~y_{\nu}{\frac{k}{M_P}} {Z\bar{N}}LH_2
\end{equation}

\noindent
where k is a coupling constant $O(1)$, $M_P$ is the Planck mass, and
$Z$ is a superfield that is invariant under the standard model gauge group.
Then the superpotential given in eqn (2) is allowed, as against the one in
eqn (1), if the various superfield have the following charge assignments 
under $Z_n$:

\begin{equation}
Z_{n}(Z)=1;~~Z_{n}(N)=1;Z_{n}(f)=0 
\end{equation}

\noindent
$f$ being any of the chiral superfields in MSSM. 
Note that (2) implies the existence of a non-renormalizable term in 
the superpotential, which, in the SUGRA context, can arise as an effective
coupling, duly suppressed by $M_P$. 

If $A_z$ and $F_z$ are respectively the vacuum expectation values (vev) of the 
scalar and auxiliary components of $Z$ (the latter being the SUSY breaking 
vev), then the Dirac mass for the neutrino is given by

\begin{equation}
m_\nu~=~{\frac{k A_z v_2}{M_P}}
\end{equation}

If $A_z$ is of the same order as the square root of the 
SUSY breaking vev $F_z$, then, in a
SUGRA scenario, $A_z~\simeq~M_X=\sqrt{m_{3/2}M_P}~\simeq~10^{11}~GeV$
(where $m_{3/2}$ is the gravitino mass),
giving $m_\nu~\simeq~10^3~eV$. This is an unacceptably large value unless one
has near-degenerate neutrinos. The solution, therefore, lies in having
$A_z~<<~\sqrt{F_z}$, i.e. in the SUSY conserving vev being much smaller 
than the 
SUSY breaking one. This can be realised, for example, in an 
O'Raifeartaigh-type model, where a hierarchy between the scalar and 
pseudoscalar components can be envisioned upon generating an effective
low-energy scalar potential for $Z$ through the condensation of some chiral
superfield in the SUSY breaking sector through non-perturbative effects:

\begin{equation}
A_z~=~16 \pi^2 k x m_{3/2}
\end{equation}

\noindent
with $x~=~O(1)$. This yields neutrino Dirac masses on the order
of $10^{-2}~eV$--the order of magnitude!

It may be relevant to comment here that $A_z$ can directly lead to small
neutrino masses even without the above mechanism in a 
gauge mediated SUSY breaking (GMSB)                    
scenario, where the gravitino is a much lighter object. In such a
case, however, mass splitting between the electron and muon neutrinos becomes
considerably smaller than what has been reported above, and can at most
place us in the solution space corresponding to vacuum oscillation. Since
the current data strongly disfavour such a solution, the GMSB option
is perhaps not of much value in this context.

There is a very similar approach \cite{ark} which puts the 
mechanism of neutrino
mass generation in a somewhat bigger perspective. It is well-known that in 
MSSM, there is no natural way to keep the Higgsino mass parameter 
$\mu$, a Supersymmetry-conserving mass, within the TeV scale. The parameter
occurs in a term $\mu H_1 H_2$ in the superpotential, and it is not clear
why it is not as big as any of the masses in the SUSY breaking sector.
On the other hand,  it is highly 
desirable to have it around the electroweak scale so that 
the minimisation condition for the scalar potential can be naturally
satisfied.

With the $\mu$-problem in view, one can think of a global symmetry group $G$
protecting the Higgs masses, and forbidding the $\mu$-term in the 
original superpotential. There can again be a gauge singlet superfield $X$
associated with the SUSY breaking sector,
transforming  non-trivially under the group $G$, which finds its way into
the superpotential via the term

\begin{equation}
W_X~=~{\frac{1}{M_P}}XH_{1}H_{2}
\end{equation}

Remember also that $F_X$, the vev auxiliary component of $X$, 
is  of the order of 
$M_X^2$ ($\simeq 10^{22}~GeV^2$) as defined above. One can immediately see that
this `SUSY breaking' vev gives rise to a $\mu$-parameter in the range of
$M_X^2/M_P~\simeq~m_{3/2}$. Thus a value of $\mu$ in the naturally expected
range is ensured.

Suppose now that the same scenario includes a right-handed neutrino superfield 
$N$. If lepton number is conserved, then one can envision a scenario where
a term in the superpotential of the form $LNH_2$ is disallowed by the charge
assignments of the corresponding superfields under $G$. However, the term  
$ENH_1$ may still be allowed if $N$, also a gauge singlet, has a different 
charge compared to the standard model superfields.
In this case, the source of the neutrino Dirac mass is

\begin{equation}
W_N~=~{\frac{1}{M_P}}XLNH_{2}
\end{equation}

Again, a scalar vev for $X$ on the order of $M_X$ leads to  
inadmissibly large neutrino masses. The interesting contribution
comes again from the auxiliary component which yields a Dirac mass,
given by

\begin{eqnarray}
m_D~\simeq~{\frac{M_X^2 v_2}{M_P^2}}~\simeq~~{\frac{v_2^2}{M_P}}
\end{eqnarray}

\noindent
which turns out to be around $10^{-3}~eV$. Once more, one is 
left with the task of preventing neutrino masses from the 
SUSY-conserving vev of $X$. This has been done in the literature
by introducing additional $U(1)$ symmetries in the SUSY breaking sector
\cite{frogniel},
and preventing, in a style similar to the one mentioned earlier,  
the scalar potential from developing a vev in the lowest order.  

Of course, the neutrino mixing pattern still needs to be explained,
the particular problem being the possibility of large mixing both
between the first and second generations and the second and third.
The only reasonable explanation of this can come from a texture
of the $XNLH_2$ coupling. However, there is no clear
understanding of how a suitable texture can be naturally ensured.

An alternative explanation of small Dirac neutrino masses has been offered 
from the assumption that the gauge singlet superfield $N$ is prevented
by a global from having an $NL H_2$ term, but that the charges are
such that a heavy superfield $H'$ can replace $H_2$ in the superpotential 
\cite{kitano}. Now, if there is mixing between $H_2$ and $H'$
after SUSY breaking, the Yukawa coupling of $N$ with the
resultant physical Higgs can have a suppression of 
${\frac{M_{EW}}{m_{H'}}}$ compared to the unsuppressed Yukawa strengths
of the corresponding charged lepton. This suppression can be used  
to account for the smallness of the Dirac neutrino masses compared to
those of their charged partners, although a nearly bimaximal
texture remains unexplained. 

Before we end this section, some remarks about radiatively generation 
Dirac masses in SUSY models are in order. Radiative generation is possible
through diagrams mediated by neutral gauginos \cite{borz1}. However, 
the fact that the right-handed neutrino superfield is 
a standard model gauge singlet
implies that such a diagram can contribute only when additional gauginos
are present. An extension of the  gauge group is therefore a 
necessity for such a mechanism to be operative.

\section{$\Delta L~=~2$ terms in the Lagrangian: Majorana masses}

If there is lepton number violation at a high scale M, it is possible to 
have $\Delta L~=~2$ neutrino mass terms via the dimension-5 operator
\cite{alta}

\begin{equation}
{\cal L}_5~=~{\frac{\lambda}{M}} LLHH
\end{equation}

\noindent
which gives neutrino masses on the order of $v^2/M$.
The most obvious model that gives rise to Majorana masses of this kind
has heavy right-handed neutrinos in the scale M, with both Yukawa couplings
with SU(2) doublets and L-violating mass terms of its own:

\begin{equation}
{\cal L}_N~=~{\frac{M}{2}}NN~+~ y_N {\bar{N}}LH
\end{equation}

\noindent
so that it is possible to generate very small neutrino mass eigenstates
without requiring inordinately small Yukawa coupling. This is the essence of 
the well-known seesaw mechanism \cite{alta,seesaw}. It is also seen that
one obtains the light neutrino masses in the expected range when $M$ is
in the Grand Unification scale of about $10^{16}~GeV$. Thus it is customary
to treat $N$ as a right-handed neutrino  belonging to the fundamental 
representation of a GUT group such as $SO(10)$. In addition, 
there can be a right-handed neutrino in each generation, so that $M$ in 
general can be a matrix, real and symmetric. The prediction
of two large mixing angles, however, is a dilemma that is yet to be
satisfactorily addressed in GUT-inspired textures of $M$.  

In what way can SUSY contribute to the Majorana mass generation mechanism
of the above type? Of course, there are numerous versions of SUSY GUT's where 
various issues related to the requisite texture have been discussed.
The recourse to SUSY GUT's has also its motivation in the observation that
the convergence of the three gauge coupling constants at high scale
is better achieved in a SUSY scenario where new threshold effects
become important around the TeV sale.
Particular SUSY  breaking schemes such as GMSB have been also invoked 
to explain the large flavour mixing necessitated by the observed data.
An important component, to which most existing studies of the subject
owe their richness, is the question of
compatibility of large mixing with the limits on lepton flavour
violating processes such as $\mu \longrightarrow e\gamma$
or $\tau \longrightarrow e\gamma$. In the SUSY
context, the mismatch between the neutrino and sneutrino mass matrices
at low-scale is a source of potentially dangerous flavour violation, and
thus the parameters of the theory must be subject to strong constraints. A 
large number of investigations in this direction can be found in the 
literature \cite{lflav}.

Here we discuss the following question: in addition to the GUT scale,
can the additional scale(s) made available to us from SUSY breaking  
be of any use in Majorana mass generation? In relation to Dirac masses,
we have already found an answer in the affirmative. Now we 
include a brief discussion related to Majorana masses \cite{borz1,ark,smaj}. 

One can, for example, extend the picture outlined in equations ()-()
by including 
a $\Delta L~=~2$ mass term for the right-handed neutrino N through
a term of the form $X^{\dagger}NN$ in the superpotential. 
In exact analogy with the situation where the $\mu$-parameter is generated
around the electroweak scale, this mass may also
be generated only from the vev of the auxiliary component of the field $X$.
The Majorana mass is thus given by

\begin{equation}
m_N~\simeq~{\frac{MX^2}{M_P}}\simeq v
\end{equation}

\noindent
v being of the same order as the electroweak scale.
For the Dirac mass, however, the scalar vev of X may be used in the
term $XLNH_2$, yielding

\begin{eqnarray}
m_D~=~{\frac{m_X v_2}{M_P}}
\end{eqnarray}

\noindent
so that the seesaw mass for the light neutrino(s) is given by
$m_D^2/m_N~\simeq~v^2/M_P$, which is in the desired range. Note that unlike
in the case of Dirac neutrinos, here one does not need to invoke a 
special mechanism such as a $U(1)$ symmetry to keep the vev of the
scalar component of $X$ small. The only thing that needs justification 
is the Majorana masses for N as well as the $\mu$-term only
out of the auxiliary component of $X$. For the latter such a condition is 
essential if one has to have $\mu$ in the electwroweak scale. It has
been argued that for the former a similar fate is expected since the 
two terms are of the same form. A contribution from the scalar vev of
$X$ would make the Majorana mass much higher and the lighter eigenvalue
much lower than is admissible, unless one can again think of a symmetry
to restrict the scalar component to a vev within the TeV scale. 
It is with an argument of this kind that the contributions from the scalar
component of $X$ have been dropped in ref. [.] from the low-energy effective 
theory, although this may not be totally above criticisms of arbitrariness.

Once more, the problem of generating two large mixing angles is not solved
in a construction of the above type. For that, one has to assume specific
textures in the XNLH couplings, which in turn requires appropriate
modelling of the SUSY breaking sector.

It has also been shown in several works \cite{ster} that 
the above principle can be
extended to include a light sterile neutrino, something that one might
require if the claims from LSND are confirmed. An additional gauge singlet 
superfield $S$ has to be added for this purpose. It is, however, necessary to
suppress the Yukawa coupling for $S$, and allow S to develop a small mass via
the scalar component of $X$, devised to be small by mechanisms
mentioned earlier. This can be ensured through an appropriate assignment 
of charge for $S$ under the group $G$.

Unlike the case of Dirac neutrinos, a Majorana neutrino can have loop-induced
masses without any extension of the gaugino sector. The second reference 
in cite{borz1} shows 
the representative diagrams from which such contributions can come.
The contribution, for which explicit expressions can be found in the 
literature, depend on the effective ${\tilde{\nu}} {\tilde{\nu}}$ 
as well as left-right mixing in the sneutrino mass matrix. 
Such loop contributions, in regions where they are substantial, may be 
required to explain (a) the mixing pattern, and (b) the mass pattern
itself where, for example, the right-handed neutrino develops a 
large Majorana mass  from the scalar component of X, making the seesaw
mechanism viable.

Before we conclude, two comments may be in order. First, the mechanisms 
discussed in this section and the last one are important,
although they might not be uniformly successful in explaining textures etc.
The reason for this is the fact that in addition to the conventional GUT
scale, here the scale $m_X$ is made available
to us by the SUSY breaking scheme. This enables one to explore newer
avenues to address the yet unanswered questions, hopefully by combining
inputs from the SUSY breaking scale with those from the GUT scale.
Secondly, the kind of models outlined here favour, among other things, 
additional right-handed sneutrinos in the electroweak scale. In fact,
Since this sneutrino mass is not restricted by the $Z$-decay width, it can
even become the lightest supersymmetric particle (LSP). This can have
considerable implications in collider phenomenology as well as issues related
to dark matter \cite{ark}.

\section{$\Delta L~=~1$ terms in the Lagrangian-- R-parity violating SUSY}

Let us next consider the case where neutrinos can acquire masses
through lepton number violating interactions at low-energy. This
is realised in R-parity violating SUSY \cite{rp1}, where
R-parity is defined as $(-)^{3B+L+2S}$. It can be seen from this definition 
that all superparticles
have $R=-1$ whereas $R$ equals $1$ for all the standard model particles.
It is also clear from above that R-parity, a multiplicatively conserved 
quantity, can be violated when B or L is violated. This makes it possible for
a superparticle to decay into two or more standard model particles, thus
rendering the LSP unstable.

In SUSY, squarks and sleptons, all spinless objects, carry lepton and baryon 
numbers. It is thus possible to violate one of these numbers
by one unit while the other is conserved. This is not possible in the
standard model due to the gauge structure and particle assignments. 
Such a provision in the SUSY scenario makes it free
from the danger of destabilising the proton. 

To see how R-parity violation actually takes place in SUSY, let us
remember that the MSSM superpotential is given by
\begin{equation}
W_{MSSM} = {\mu} {H}_1 {H}_2 + h_{ij}^l {L}_i {H}_1 {E}_j^c
+ h_{ij}^d {Q}_i {H}_1 {D}_j^c + h_{ij}^u {Q}_i
{H}_2 {U}_j^c
\end{equation}
where the last three terms give the Yukawa interactions corresponding to
the masses of the charged leptons and the down-and up-type quarks, and
$\mu$ is the Higgsino mass parameter.

When R-parity is violated, the following additional terms can be added to
the superpotential \cite{rpsup}: 
\begin{equation}
W_{\not R} = \lambda_{ijk} {L}_i {L}_j {E}_k^c +
\lambda_{ijk}' {L}_i {Q}_j {D}_k^c +
\lambda_{ijk}''{U}_i^c {D}_j^c {D}_k^c + \epsilon_i {L}_i {H}_2
\end{equation}
with the $\lambda''$-terms causing B-violation, and the remaining
ones, L-violation. The need to avoid proton decay usually prompts
one to have {\em only one} of the two types of 
nonconservation at a time. Since we are concerned with neutrino masses 
here, we will consider only lepton number violating effects.

The $\lambda$-and $\lambda'$-terms have been 
widely studied in connection  with various 
phenomenological consequences, enabling one to impose various 
kinds of limits on them \cite{limits}. Their contributions to 
neutrino masses can be only through loops, and 
their multitude (there are 36 such couplings
altogether) makes the necessary adjustments possible for reproducing
the requisite values of neutrino masses and mixing angles. We shall come 
back to these `trilinear' effects later.

More interesting, however, are the three bilinear terms 
$\epsilon_{i}L_{i}H_2$ \cite{bilerv}. Since there are only three terms 
of this type, the model looks simpler and more predictive with them alone 
as sources of R-parity violation. This is particularly so because
the physical effects of the trilinear terms can be generated from the
bilinears by going to the appropriate bases \cite{soubm}. In addition, 
they  have interesting consequences of their own \cite{morebil}, 
since terms of the type 
$\epsilon_{i}L_{i}H_2$ imply mixing between the Higgsinos and the 
charged leptons and neutrinos. In  this discussion, we shall assume,
without any loss of generality,
the existence of such terms involving only the second and third 
families of leptons.  

The scalar potential in such a case contains the following
terms which are bilinear in the scalar fields:
\begin{eqnarray}
V_{\rm scal} &=& m^2_{L_3} |{\tilde L}_3|^2 + m^2_{L_2} |{\tilde L}_2|^2 + 
m^2_1 |H_1|^2 + m^2_2 |H_2|^2 + B \mu H_1 H_2 \nonumber\\*
&& + B_2 \epsilon_2 {\tilde L}_2 H_2 +B_3 \epsilon_3 {\tilde L}_3 H_2 
+ \mu \epsilon_3 {\tilde L}_3^{*} H_1 + \mu \epsilon_2 {\tilde L}_2^{8} H_1 
+ .....
\end{eqnarray} 
where $m_{L_i}$  denotes the mass of the {\it i}th scalar doublet
at the electroweak scale, and $m_1$ and $m_2$ are the mass parameters
corresponding to the two Higgs doublets. $B$, $B_2$ and $B_3$ are  
soft SUSY-breaking parameters.

An immediate consequence of the additional (L-violating) soft terms in
the potential is a set of non-vanishing vacuum expectation values
(vev) for the sneutrinos \cite{9}. 
This gives rise to the mixing of electroweak gauginos with neutrinos (and
charged leptons) through the sneutrino-neutrino-neutralino (and
sneutrino-charged lepton-chargino) interaction
terms. The hitherto massless neutrino states enter into the neutralino 
mass matrix through such mixing and acquire
see-saw masses, where the high scale is supplied by the massive states. 
massive states. The parameters controlling the neutrino sector in
particular and R-parity violating effects in general are the 
bilinear coefficients $\epsilon_2$ , $\epsilon_3$  and the  
soft parameters $B_2$, $B_3$.

For a better understanding, let us perform a basis rotation and
remove the R-parity violating bilinear terms from the superpotential 
by  suitably redefining
the lepton and Higgs superfields. This, however, does not eliminate
the effects of these terms, since they now take refuge in the
scalar potential. The sneutrino vev's in this rotated basis (which are
functions of both and the $\epsilon$'s and the soft terms in the
original basis) trigger neutrino-neutralino
mixing. Consequently, the $6 \times 6$ neutralino mass matrix in this
basis has the following form:

\begin{equation}
{\cal M} =  \left( \begin{array}{cccccc}
  0 & -\mu & \frac {gv} {\sqrt{2}} & 
  -\frac {g'v} {\sqrt{2}} & 0 & 0 \\
  -\mu & 0 & -\frac {gv'} {\sqrt{2}} 
       & \frac {g'v'} {\sqrt{2}} & 0 & 0 \\
 \frac {gv} {\sqrt{2}} & -\frac {gv'} {\sqrt{2}} & M & 0 & -\frac {gv_3} 
 {\sqrt{2}} & -\frac {gv_2} {\sqrt{2}} \\
 -\frac {g'v} {\sqrt{2}} & \frac {g'v'} {\sqrt{2}} & 0 & M' & 
  \frac {g'v_3} {\sqrt {2}} & \frac {g'v_2} {\sqrt {2}} \\
 0 & 0 & -\frac {gv_3} {\sqrt {2}} & \frac {g'v_3} {\sqrt {2}} & 
 0 & 0  \\
 0 & 0 & -\frac {gv_2} {\sqrt {2}} & \frac {g'v_2} {\sqrt {2}} & 
 0 & 0 
 \end{array}  \right)    
\end{equation}                                 
where the successive rows and columns correspond to
(${\tilde H}_2, {\tilde H}_1, -i\tilde{W_3}, 
-i\tilde{B}, \nu_\tau, \nu_\mu$),   $\nu_\tau$ and  $\nu_\mu$ being the
neutrino flavour eigenstates in this basis. Also,  with the sneutrino
vev's denoted by $v_2$ and $v_3$, 
$$
v\ \ (v') = \sqrt{2}\ {\left(\frac {m^2_Z} {\bar{g}^2} 
 - \frac {v^2_2+v_3^2} {2} \right)}
^{\frac {1} {2}} {{\sin} \beta}\ \ ({{\cos} \beta})
$$
$M$ and $M'$ being the ${\rm SU(2)}$ and ${\rm U(1)}$ gaugino mass
parameters respectively, and $\bar{g}=\sqrt{g^2+{g'}^2}$.

One can now define two states $\nu_3$ and $\nu_2$, where
\begin{equation}
\nu_3 =\cos \theta\ \nu_\tau  + \sin \theta\ \nu_\mu 
\end{equation}
and $\nu_2$ is the orthogonal combination, the neutrino mixing angle
being given by
\begin{equation}
\cos \theta = \frac{v_3}{\sqrt{v^2_2 + v^2_3}}
\end{equation}                  
Clearly, the state $\nu_3$ --- which alone develops cross-terms with
the massive gaugino states --- develops a see-saw type mass at the
tree-level. The orthogonal combination $\nu_2$ still remains massless.
Interestingly, now we have a new seesaw mechanism, where the SUSY
breaking scale in the observable sector takes the place of the GUT
scale or the scale $m_X$ discussed in the earlier sections.

The massive state $\nu_3$ can be naturally used to account for
atmospheric neutrino oscillations, with $\Delta m^2 = m_{\nu_3}^2.$
Large angle mixing between the $\nu_\mu$ and the $\nu_\tau$
corresponds to the situation where $v_2 \simeq v_3$.

The tree-level mass here is clearly controlled by 
$v'~=~\sqrt{v_2^2 + v_3^2}$. This quantity, defined as the `effective'
sneutrino vev in the basis where the $\epsilon$'s are rotated away,
can be treated as a basis-independent measure of R-parity violation in
such theories \cite{sacha}. The SK data on 
atmospheric neutrinos restrict $v'$ to
be on the order of a few hundred keV's.  However, it should
be remembered that $v'$ is a function of $\epsilon_2$ and $\epsilon_3$,
both of which can still be as large as on the order of the electroweak
scale. For example, in SUGRA-based models, 
it is possible to have a very small value of $v'$
starting from large $\epsilon$'s, provided that one assumes the
R-conserving and R-violating soft terms (as also the slepton and $Y=1$
Higgs mass parameters) to be the same at the
scale of dynamical SUSY breaking at a high energy \cite{sugbil}.

Also, one has to address the question as to whether the treatment of
$\nu_3$ and $\nu_2$ as  mass eigenstates is proper,
from the viewpoint of the charged lepton mass matrix being diagonal
in the basis used above. In fact, it can be shown that this is strictly
possible when $\epsilon_2$ is much smaller than $\epsilon_3$, failing
which one has to give a further basis rotation to define the
neutrino mass eigenstates. However, the observable consequences
described here are still valid, with the requirement of near-maximality
shifted from the angle $\theta$ to the effective mixing angle.

Furthermore, a close examination of the scalar potential in such a
scenario reveals the possibility of additional mixing among the
charged sleptons, whereby flavour-changing neutral currents (FCNC) can
be enhanced. It has been concluded after a detailed study 
that the suppression of FCNC requires one to have the
$\epsilon$-parameters to be small compared to the MSSM parameter $\mu$
(or, in other words, to the electroweak scale) unless there is a
hierarchy between $\epsilon_2$ and $\epsilon_3$ \cite{asesh1}.

However, one still needs to find a mechanism for mass-splitting
between the massless state $\nu_2$ and the electron neutrino, and
to explain the solar neutrino puzzle. 
While there exist studies
which attempt to explain both the puzzles in terms of bilinear terms only,
the existence of the various $\lambda$ and $\lambda'$-terms can also 
give rise to loop contributions to the neutrino mass matrix \cite{loopm}.

The generic expression
for such loop-induced masses is
\begin{eqnarray}
(m^{\rm loop}_\nu)_{ij} &\simeq& \frac {3} {8\pi^2}  m^d_k m^d_p M_{SUSY} 
\frac {1} {m^2_{\tilde q}} {\lambda_{ikp}'\lambda_{jpk}'}  
\nonumber\\* && + 
\frac {1} {8\pi^2}  m^l_k m^l_p M_{SUSY} 
\frac {1} {m^2_{\tilde l}} {\lambda_{ikp}\lambda_{jpk}}  
\end{eqnarray}         
where $m^{d,(l)}$ denote the down-type quark (charged lepton) masses.
${m^2_{\tilde l}}$, ${m^2_{\tilde q}}$ are the slepton and 
squark mass squared. $M_{SUSY}(\sim \mu)$ is the effective scale of 
supersymmetry breaking. The mass eigenvalues
can be obtained by including these loop contributions
in the mass matrix.

Again, it should be noted that there may be other ways of looking at the 
problem. For example, it has been shown in \cite{anjans} that, if one assumes
either purely bilinear or purely trilinear R-violating 
interactions at a high scale, running of the
mass parameters can lead to significant sneutrino vev's at low energy,
and at the same time generate loop-induced masses.

If we want the mass thus induced for the second generation neutrino to
be the right one to solve the solar neutrino problem, then one obtains
some constraint on the value of the $\lambda'$s as well as $\lambda$s.
In order to generate a splitting between the two residual massless
neutrinos, $\delta m^2 \simeq 5 \times 10^{-6}\ {\rm {eV^2}}$ (which
is suggested for an MSW solution ), a SUSY breaking mass of about 500
GeV implies $\lambda'\ (\lambda) \sim 10^{-4}~-~10^{-5}$.  

An interesting aspect of the scenario described above is that
it can have distinctive signatures in collider experiments.
The most striking ones among them pertain to decays of the
lightest neutralino, produced either directly or via cascades. 
In presence of only the trilinear R-violating terms in the
superpotential, the lightest neutralino can have various three-body
decay modes which can be generically described by $\chi^{0}
\longrightarrow \nu f \bar{f}$ and $\chi^{0} \longrightarrow l f_1
\bar{f_2}$, $f$, $f_1$ and $f_2$ being different quark and lepton
flavours that are kinematically allowed in the final state.

Due to the mixing between neutrinos and neutralinos as also between charged
leptons and charginos, the bilinears open up additional decay channels for the
lightest neutralino, namely, $\chi^{0} \longrightarrow l W$ and
$\chi^{0} \longrightarrow \nu Z$.  When the neutralino is heavier than
at least the W, these two-body channels dominate over three-body
ones over a large region of the parameter space, the effect of which
can be observed in colliders such as the upgraded Tevatron, the LHC
and a proposed high-energy electron-positron collider. In addition, 
superparticles such as the stop can sometimes decay dominantly via R-parity
violating interactions, thereby altering the observed signals. Different
observable quantities related to these decays have been studied in 
recent times \cite{visbil,porod,colsig,tevsig}.

Here we would like to stress upon one distinctive feature of the
scenario that purportedly explains the SK results with the help of
bilinear R-parity violating terms. It has been found that over almost
the entire allowed range of the parameter space in this connection,
the lightest neutralino is dominated by th Bino.  A glance at the
neutralino mass matrix reveals that decays of the neutralino ($\simeq$
Bino) in such a case should be determined by the coupling of different
candidate fermionic fields in the final state with the massive
neutrino field $\nu_3$ which has a cross-term with the Bino. Large
angle neutrino mixing, on the the hand, implies that $\nu_3$ should
have comparable strengths of coupling with the muon and the tau. Thus,
a necessary consequence of the above type of explanation of the SK
results should be comparable numbers of muons and tau's emerging from
decays of the lightest neutralino, together with a $W$-boson in each
case \cite{visbil,porod}.

Of course, the event rates in the channel mentioned above will depend on
whether the two-body decays mentioned above indeed dominate over the
three-body decays. The latter are controlled by the size of the
$\lambda$-and $\lambda'$-parameters. If
these parameters have to be of the right size 
to explain the mass-splitting required by the solar neutrino
deficit, then, for large angle MSW case, the decay widths driven by
the trilinear term are smaller than those for the two-body decays by at 
least an order of magnitude.

The other important consequence of this picture is a large
decay length for the lightest neutralino. We have already mentioned
that the atmospheric neutrino results restrict the basis-independent
R-violating parameter $v'$ to the rather small value of a few hundred
keV's. This value affects the mixing angle involved in calculating the
decay width of the neutralino, which in turn is given by the formula
\begin{equation}
L =  \frac{\hbar}{\Gamma} 
\times 
\frac{p}{M(\tilde{\chi}^0_1)} 
\end{equation}  
where $\Gamma$ is the rest frame decay width of 
the lightest neutralino and $p$, its momentum.  The
decay length decreases for higher neutrino masses, as a result of the
enhanced flipping probability between the Bino and a neutrino, when
the LSP is dominated by the Bino. Also, a relatively massive
neutralino decays faster and hence has a smaller decay length. The
interesting fact here is that even for a neutralino as massive as 250
GeV, the decay length is as large as about 0.1 to 10 millimetres, which
should be observable in a detector \cite{visbil}.

If the lightest neutralino can have two-body charged current decays,  
then the Majorana character of the latter also leads to the possibility 
of like-sign dimuons and ditaus from pair-produced neutralinos \cite{tevsig}. 
Modulo the efficiency of simultaneous identification of W-pairs, these
like-sign dileptons can also be quite useful in verifying the type
of theory discussed here.
 
\section{Some other possibilities}

\subsection{Nearly degenerate neutrinos}

If the mass ranges to which the neutrino eigenstates belong are represented
by mass-squared differences indicated by the solar and atmospheric neutrino
deficits, then it is difficult to account for the hot dark matter content
of the universe in terms of neutrinos. A way to surmount the difficulty
is to postulate nearly degenerate neutrinos \cite{visdeg}. Degeneracy 
also helps us understand large mixing in a somewhat `natural' manner. 
At the same time, with a sterile neutrino with
mass in the similar order, it may provide an explanation of the
LSND results if they are substantiated. 

However, there are problems with degenerate neutrinos. The limit on
the electron neutrino mass from tritium beta decay provides the
first restriction. More seriously,
if neutrinos are of Majorana character, then degeneracy can come into 
serious conflicts with constraints imposed from the search for
neutrinoless double-beta decay. There have been efforts to circumvent this
difficulty by proposing neutrino mixing matrices which effect a cancellation
between different eigenstates in such decay \cite{thap}. Also, the
literature contains proposals of a partial lifting of
the degeneracy. On the whole, these scenarios cannot be completely ruled out, 
though some natural foundation for any of the models is yet to be found.

In the context of SUSY, too, effort have been on to justify degenerate
neutrino scenarios, and we shall mention only one approach here \cite{casas}.
In this work, the close degeneracy of the neutrino masses can be
{\it a~priori}  postulated
to come from the form of the neutrino mass matrix at the Planck scale. 
Following works, for example of Georgi and Glashow, the matrix 
can be taken to correspond  exactly to bimaximal mixing at the
Planck scale. The evolution of the mass parameters should provide the
requisite splittings at low energy. The evolution is crucially controlled by
Yukawa couplings, and this is where the dependence on $\tan~\beta$, the
ratio of the vev's of the two Higgs doublets becomes most important. However,
it has been shown \cite{casas} that the solution 
space corresponding to the large mixing angle (LMA) MSW 
mechanism yields an inadmissible mass splitting unless $\tan~\beta$
is very small, which is again incompatible with accelerator data. On the
other hand a seesaw approach, with a high-scale Majorana mass in the
range of $10^{10}~GeV$, leads to acceptable MSW solutions in the LMA regions.
This, however, gives the best fit for $\tan~\beta~\simeq~2$ which is
at the very edge of the phenomenologically viable MSSM parameter space.

\subsection{Neutrino mass from unusual SUSY breaking terms}

We normally agree to have `soft' SUSY breaking terms only, the main reason
being the need to control quadratic divergence of scalar masses. However,
since the SUSY breaking interaction is usually an effective theory,
one may expect higher order terms also to creep into the picture. Though
such `hard' terms are potential threats to the stability of scalar masses,
they are suppressed by some power(s) of the cut-off scale for the effective
theory, which in this case turns out to be the Planck mass $m_P$. Thus
the quadratic corrections effectively shift the scalar masses by very small
amounts, and the hard terms are usually ignored as phenomenologically
insignificant. Such a possibility is conceivable also in the schemes 
suggested in reference [ ], with an enlarged SUSY breaking sector. Also,
such terms have sometimes been exploited 
to stabilise flat directions of the scalar 
potential and generate intermediate scale vev's.

It has been suggested \cite{frere} that some 
of these suppressed higher-dimensional terms 
may be responsible for neutrino masses. This is true in particular if lepton
number is violated. Under such circumstances, one may, for example, have a 
gauge invariant term in the Lagrangian, of the form

\begin{equation}
{\cal L}_{hard}~=~h(\epsilon_{ij} {\tilde{L_i}}H_{2j})^2
\end{equation}

\noindent
where $\epsilon_{ij}$ is the completely antisymmetric rank-2 tensor.
The dimensionless coupling $h$ in this case depends on
$(m_X^2/M_P^2)^n$ where $n$ depends on the specific SUSY breaking mechanism. 
Note that this term is L-violating but R-parity conserving.

Such a term generates Majorana neutrino masses at one-loop level, involving
virtual sneutrinos and $SU(2)$ gauginos. The induced mass has been shown
to be of the form

\begin{equation}
m_\nu~\simeq~ {\frac{h g^2 v_2^2}{32 \pi^2 m_{\tilde{\nu}}}} 
F(M_2^2/m_{\tilde{\nu}}^2)
\end{equation}

\noindent
where $M_2$ and $m_{\tilde{\nu}}$ are the $SU(2)$ gaugino and sneutrino masses
respectively. The function F ranges between $0.5$ and $0.1$ for 
phenomenologically allowed values of the mass ratio in the argument.

Using such an expression, it can be seen that for a sneutrino mass
in the range of $100~GeV$ and phenomenologically allowed values of
the ratio of the Higgs vev's, the induced neutrino mass turns out to be
too small to be consistent with observed results if $n~=~1$, while
for $n~=~1/2$ it stays a little above the acceptable range. \footnote{In 
reference \cite{frere}, the net induced mass has been claimed to be on the
order of $M_X^2/M_P^3$ assuming that $n=1$, which is misleading, for,
as has been subsequently admitted in the same paper, the factor
of $32\pi^2$ makes the contribution `somewhat smaller'.} A mechanism
of the above kind therefore favours SUSY breaking schemes where 
the dimension-4 terms shown in equation (4) are suppressed by some
{\em fractional power} of the ratio $(m_X^2/M_P^2)$. An additional problem,
of course, is to explain neutrino mixing in this scheme, for which
the evolution of the term shown in equation () to low energies has 
to play a role.

\section {Concluding remarks}
I have reviewed some of the various ways in which a SUSY scenario can be
responsible for the generation of neutrino masses. I must admit that 
there are many interesting approaches left out in this review. The point 
which has been emphasised here is the fact that SUSY notionally
brings in additional mass scales into low-energy physics, which can have
a role to play in the domain of neutrinos. Also, some special status of the 
right-handed neutrino superfield with respect to the governing symmetry
in the SUSY breaking sector might well be responsible for the different
nature of neutrino masses with respect to those of the other fermions. 
Such a point of view can be applied to both Dirac and Majorana masses,
and also to cases which give rise to light sterile neutrinos. Side by side,
the $\Delta L=1$ terms in the superpotential of
an R-parity breaking SUSY theory can use the electroweak symmetry breaking
scale itself in a spectacular manner to explain not only neutrino masses but
also their mixing pattern. Several of the theories discussed
above have implications in other aspects of electroweak phenomenology including
high-energy collider phenomena, which, quite desirably, integrates
neutrino-related model-building into a much bigger canvas. Scenarios with 
degenerate neutrinos  can also be encompassed by SUSY models. And finally, 
there exists the  interesting conjecture that the otherwise undesirable 
hard SUSY breaking terms, suppressed by some power(s) of the Planck mass, 
can after all have a role to play in neutrino physics. 

It should be admitted finally that flavour mixing, especially that of the
bimaximal type, still requires special model assumptions. A better 
understanding of SUSY breaking schemes is necessary for further insight into
the matter. 

{\bf Acknowledgements:} This work was partially supported by the Board of Research in
Nuclear Sciences, Government of India.

\end{document}